\documentclass{ws-ijmpa}

\begin{document}
\markboth{B.A. Arbuzov and I.V. Zaitsev}
{On a possibility to calculate fundamental parameters of the SM
}
%
%

\title{On a possibility to calculate fundamental parameters of
the Standard Model}

\author{Boris A. Arbuzov }

\address{D.V. Skobeltsyn Institute for Nuclear Physics of M.V.
Lomonosov Moscow State University\\ Leninskie gory 1, 119991 Moscow,
Russia\\
arbuzov@theory.sinp.msu.ru}
\author{Ivan V. Zaitsev}
\address{D.V. Skobeltsyn Institute for Nuclear Physics of M.V.
Lomonosov Moscow State University\\ Leninskie gory 1, 119991 Moscow,
Russia\\
zaitsev@theory.sinp.msu.ru}

\maketitle
\date{\today}
\begin{abstract}
The problem of a calculation of
parameters of the Standard Model is considered in the framework of the
compensation approach. Conditions for a spontaneous generation of effective
interactions of fundamental fields are shown to lead to
sets of equations for parameters of a theory.
A principal possibility to calculate mass ratios of fundamental
quarks and leptons is demonstrated, as well of
mixing angles of quarks, {\it e.g.} of the Cabibbo angle.
A possibility of a spontaneous generation of an effective
interaction of electroweak gauge bosons $W^a$ and $B$ is demonstrated.
In case of a realization of a non-trivial solution of a set of compensation
equations, parameter $\sin^2\theta_W$ is defined. The non-trivial solution
 is demonstrated to provide a satisfactory value for
the electromagnetic fine structure constant $\alpha$ at scale $M_Z$: 
$\alpha(M_Z) = 0.00772$. The results being obtained may be considered
as sound arguments on behalf of a
possibility of a calculation of parameters of the Standard Model.


\end{abstract}

\ccode{PACS numbers: 11.15.Tk; 12.15.-y; 12.15.Ji; 12.90.+b; 14.70.Fm}

\maketitle
\section{Introduction}
In works~\cite{BAA04}\cdash\cite{AZ20132},
N.N. Bogoliubov compensation principle~\cite{Bog1,Bog2}
was applied to studies of a spontaneous generation of effective non-local interactions in renormalizable gauge theories. The method and applications
are also
described in full in the    book~\cite{ABOOK}.

In particular, papers~\cite{BAA09}\cdash\cite{AVZ2}  deal with an application of the approach to the electro-weak interaction and a possibility of spontaneous generation of effective anomalous three-boson interaction of the form
\begin{eqnarray}
& &-\,\frac{G}{3!}\,F\,\epsilon_{abc}\,W_{\mu\nu}^a\,W_{\nu\rho}^b\,
W_{\rho\mu}^c\,;
\label{FFF}\\
& &W_{\mu\nu}^a\,=\,
\partial_\mu W_\nu^a - \partial_\nu W_\mu^a\,+g\,\epsilon_{abc}
W_\mu^b W_\nu^c\,.\nonumber
\end{eqnarray}
with uniquely defined form-factor $F(p_i)$, which guarantees effective interaction~(\ref{FFF}) acting in a limited region of the momentum space. It was done in the framework of an approximate scheme, which accuracy was estimated to be $\simeq (10 - 15)\%$~\cite{BAA04}. Would-be existence of effective interaction~(\ref{FFF}) leads to important non-perturbative effects in the electro-weak interaction. It is
usually called anomalous three-boson interaction and it is considered for long time on phenomenological grounds~\cite{Hag1,Hag2}. Our interaction constant $G$ is connected with
conventional definitions in the following way
\begin{equation}
G\,=\,-\,\frac{g\,\lambda}{M_W^2}\,;\label{Glam}
\end{equation}
where $g \simeq 0.65$ is the electro-weak coupling.
The best limitations for parameter $\lambda$ read~\cite{PDG}
\begin{equation}
\lambda_\gamma = -\,0.022\pm0.019\,;\quad
 \lambda_Z = -\,0.09\pm0.06\,; \label{lambda1}
\end{equation}
where subscript denote a neutral boson being involved in the experimental definition of $\lambda$.

\begin{figure}
\includegraphics[scale=0.9,width=12cm]{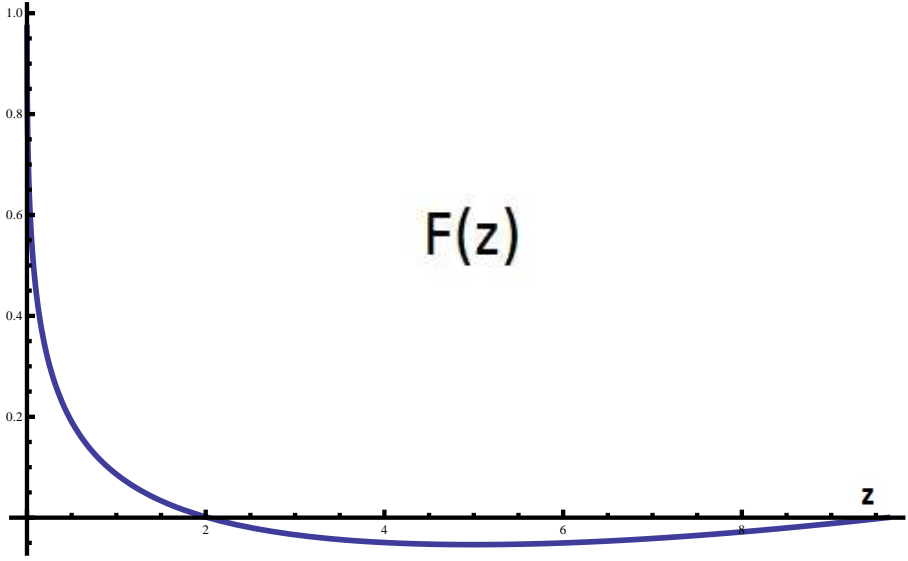}
\caption{The behavior of the form-factor for the electro-weak theory.}
\label{fig:compenG}
\end{figure}

Solution of the analogous compensation procedure in QCD correspond to  $g(z_0)=3.8$~\cite{AZ20132}. For the electro-weak interaction we have~\cite{AZ11,AVZ2}
\begin{equation}
g(z_0) = 0.60366\,;\quad z_0 = 9.6175\,;\quad |\lambda| = 2.88\cdot 10^{-6}\,.\label{eq:gz0}
\end{equation}
Here $z_0$ is a dimensionless parameter, which is connected with
value of a boundary momentum, that is with effective cut-off
$\Lambda$ according  to the following definition~\cite{AZ11,AVZ2}
\begin{equation}
\frac{2\,G^2\,\Lambda^4}{1024\,\pi^2}\,=\,
\frac{g^2\,\lambda^2\,\Lambda^4}{512\,\pi^2\,M_W^4}\,=\, z_0 \,.\label{eq:Lambda}
\end{equation}
It is instructive to present in Fig.~\ref{fig:compenG} the behavior of form-factor $F(p,-p,0)$ in dependence on momentum $p$, where
\begin{equation}
z\,=\,\frac{G^2\,p^4}{512\,\pi^2}\,;\label{eq:zdef}
\end{equation}
and $F(z)\,=\,0$ for $z\,>\,z_0$.
As a rule the existence of a non-trivial solution of a compensation
equation impose essential restrictions on parameters of a problem. Just
the example of these restrictions is the definition of coupling constant
$g(z_0)$ in~(\ref{eq:gz0}). It is advisable to consider other possibilities
for spontaneous generation of effective interactions and to find out, which restrictions on physical parameters may be imposed by an existence of non-trivial solutions. In the present work we consider possibilities of definition of important physical parameters: mixing angles and mass
ratios of
elementary constituents of the Standard Model.

\section{A model for mass relations of quarks and leptons}

Following the approach used in works~\cite{BAA04}\cdash\cite{AZ20132} let us formulate the compensation equations for
would-be four-fermion interaction of two types of quarks
and two leptons, that is we consider one generation of
fundamental fermions. For the simplicity we
call them $"u"$, $"d"$, $"e"$ and $"\nu"$,
which in the standard way are represented by their left $\psi_L$ and
right $\psi_R$ components. We admit initial masses for all participating fermions to be zero and we will look for possibility of them to acquire
masses $m_i, i = 1,...4$ respectively due to interaction with scalar
Higgs-like composite field.

Then let us consider a  possibility of spontaneous generation 
of the following interaction, which is constructed by close analogy with the 
well-known Nambu -- Jona-Lasinio effective interaction~\cite{Nambu}\cdash\cite{VolRad}
\begin{eqnarray}
& &L_{eff}^{F}\,=\,G_1 \bar u_L\,u_R\,\bar u_R\,u_L  +
G_2 \bar d_L\,d_R\,\bar d_R\,d_L  + G_4 \bar e_L\,e_R\,\bar e_R\,e_L + \nonumber\\
& &G_3(\bar u_L\,u_R\,\bar d_R\,d_L + \bar d_L\,d_R\,\bar u_R\,u_L) +
G_5(\bar u_L\,u_R\,\bar e_R\,e_L + \bar e_L\,e_R\,\bar u_R\,u_L)+\label{eq:Luuddee}\\
& &G_6(\bar e_L\,e_R\,\bar u_R\,u_L+\bar e_R\,e_L\,\bar u_L\,u_R)+G_7\,\bar \nu_L\,\nu_R\,
\bar \nu_R\,\nu_L +\nonumber\\
& &G_8(\bar \nu_L\,\nu_R\,\bar d_R\,d_L + \bar d_L\,d_R\,\bar \nu_R\,\nu_L)+G_9(\bar \nu_L\,\nu_R\,\bar u_R\,u_L + \bar u_L\,u_R\,\bar \nu_R\,\nu_L)+\nonumber\\
& &G_{10}(\bar \nu_L\,\nu_R\,\bar e_R\,e_L + \bar e_L\,e_R\,\bar \nu_R\,\nu_L).\nonumber
\end{eqnarray}
Here all coupling constants $G_i$ have dimension of the inverse mass
squared $M^{-2}$.

Now we would like to find out, if the four-fermion
interaction~(\ref{eq:uuddee})
could be spontaneously generated.
In doing this we again proceed with the
\index{add-subtract}add-subtract procedure
\begin{eqnarray}
& &L = L_0\,+\,L_{int}\,;\quad L_{int}\,=\,L_{0int}\,+\,L_{eff}^{F}\,;\nonumber\\
& &L_0 = \sum_{u,d}\bar q(x)(\imath\partial_\alpha\gamma_\alpha - m)q(x)
+
\sum_{e,\nu}\bar l(x)(\imath\partial_\alpha\gamma_\alpha - m)l(x) - L_{eff}^{F}\,; \label{eq:Leff}
\end{eqnarray}
here $L_{0int}$ is an initial interaction Lagrangian.
Then we have to compensate the undesirable term $L_{eff}$ in the newly
defined free Lagrangian. The relation, which serve to accomplish this
goal, is called compensation equation. Necessarily we use approximate form of this equation.
In diagram form the compensation equation
for four fermions participating the
interaction in one-loop approximation is presented in Fig.~{\ref{fig:6equat}.
\begin{figure}
\includegraphics[scale=0.85,width=11.2cm]{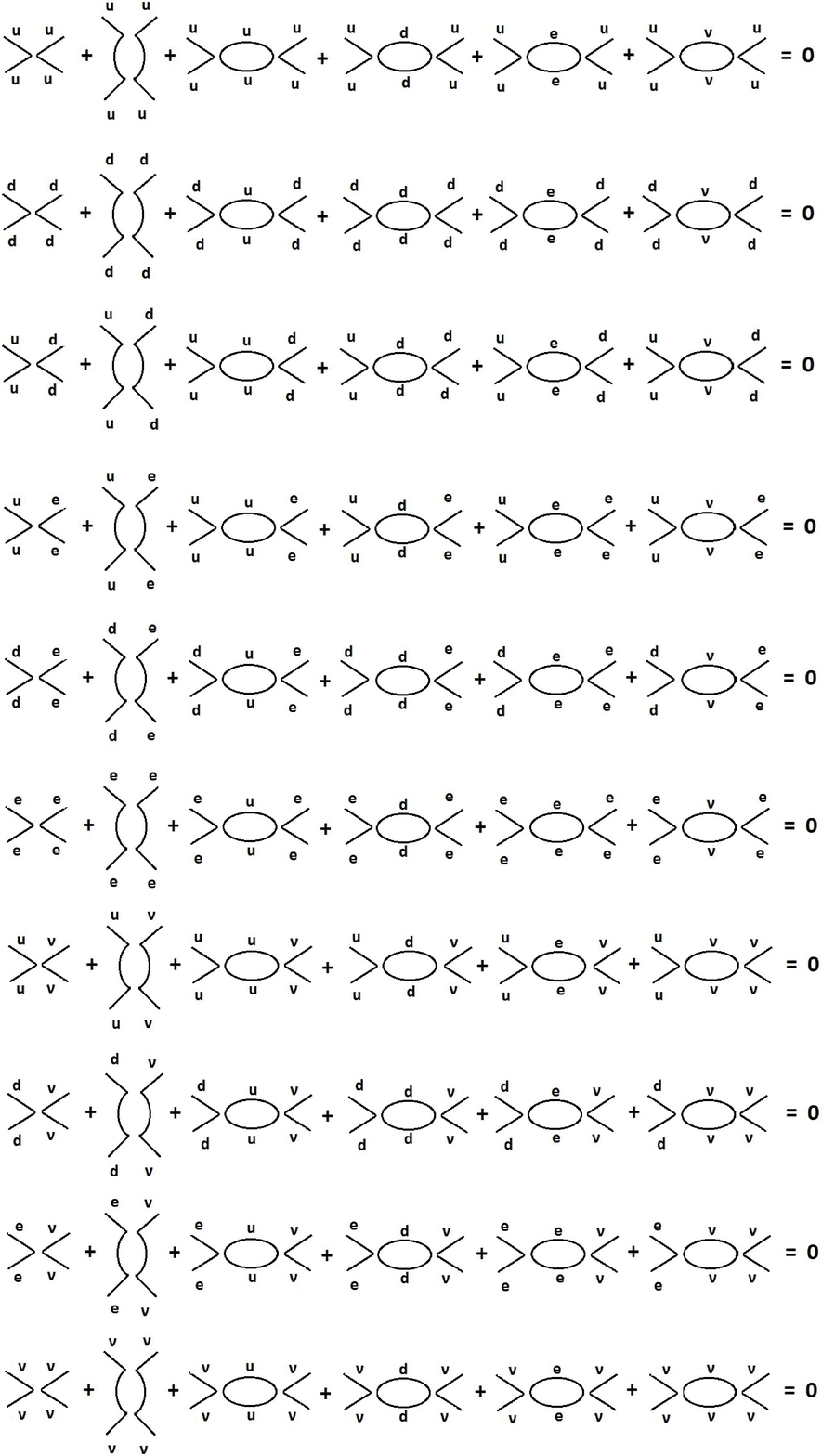}
\caption{Diagram representation of the compensation
equation for spontaneous generation of interaction~(\ref{eq:uuddee}).
Notations of quarks and lepton are shown by corresponding lines.}
\label{fig:6equat}
\end{figure}

Let us define effective cut-off $\Lambda$ in integrals of
equation~(\ref{eq:uuddee}). We shall see below, that $\Lambda$ may be defined in the course of solution of compensation equations. With account of this definition we introduce the following dimensionless variables
\begin{eqnarray}
& &y_1\,=\,\frac{G_1\,\Lambda^2}{8 \,\pi^2}\,;\;y_2\,=\,\frac{G_2\,\Lambda^2}{8 \,\pi^2}\,;\;y_3\,=\,\frac{G_3\,\Lambda^2}{8 \,\pi^2}\,;\nonumber\\
& &z_1\,=\,\frac{G_4\,\Lambda^2}{8 \,\pi^2}\,;\;z_2\,=\,\frac{G_7\,\Lambda^2}{8 \,\pi^2}\,;\;z_3\,=\,\frac{G_{10}\,\Lambda^2}{8 \,\pi^2}\,;
\label{eq:GY8}\\
& &x_1\,=\,\frac{G_5\,\Lambda^2}{8 \,\pi^2}\,;\;x_2\,=\,\frac{G_9\,\Lambda^2}{8 \,\pi^2}\,;\;
x_3\,=\,\frac{G_6\,\Lambda^2}{8 \,\pi^2}\,;\nonumber\\
& &x_4\,=\,\frac{G_8\,\Lambda^2}{8 \,\pi^2}\,;\;\nonumber\\
& &\xi_1\,=\,\frac{m_2}{m_1}\,;
\;\xi_2\,=\,\frac{m_3}{m_1}\,;
\;\xi_3\,=\,\frac{m_4}{m_1}.\nonumber
\end{eqnarray}

Then we consider scalar bound state consisting of all possible
fermion-antifermion combinations $\bar u u$,
$\bar d d$, $\bar e e$ and $\bar \nu \nu$. The corresponding set of Bethe-Salpeter
equations is shown in Fig.~\ref{fig:3equat}.
In this way we come to the following set of ten compensation equations presented in
 Fig.~\ref{fig:6equat} and four
Bethe-Salpeter equations shown in Fig.~\ref{fig:3equat}. Let us note,
that in Fig.~\ref{fig:3equat} we present also wouldbe contributions of
gauge bosons exchanges, which in the calculations of the present section
are not taken into account.
Note also, that terms with factor $A$ arise from vertical diagrams in Fig.~{\ref{fig:6equat}. Let us remind, that the sign minus before linear
terms in compensation equations is connected with opposite signs of terms corresponding to effective interactions in the new free Lagrangian and in the new interaction Lagrangian.

\begin{eqnarray}
& &-y_1\,+ \,A\,y_1^2\,+\,3(y_1^2+y_3^2)\,+\,x_1^2\,+\,x_2^2\,=0\,;\nonumber\\
& &-y_2\,+ \,A\,y_2^2\,\xi_1^2\,+\,3(y_2^2+y_3^2)\,+\,x_3^2\,+x_4^2\,=0\,;\nonumber\\
& &-y_3\,+ \,A\,y_3^2\,\xi_1\,+\,3\,y_3(y_1+y_2)\,+\,x_1\,x_3\,+\,x_2 x_4\, =0\,;\nonumber\\
& &-z_1\,+ \,A\,z_1^2\, \xi_2^2\,+\,3\,(x_1^2+x_3^2)\,+z_1^2\,+\,z_3^2\,=0\,;\nonumber\\
& &-z_2\,+ \,A\,z_2^2\,\xi_3^2\,+\,3\,
(x_2^2+x_4^2)\,+z_2^2\,+\,z_3^2\,=0\,;\nonumber\\
& &-z_3\,+ \,A\,z_3^2\,\xi_2\,\xi_3\,+\,3\,
(x_1\,x_2+x_3\,x_4)\,+\,z_1 z_3\,+\,z_2 z_3\,=0\,;\nonumber\\
& &-x_1\,+ \,A\,x_1^2\,\xi_2\,+\,3(x_1 y_1+x_3 y_3)\,+\,
x_1 z_1\,+\,x_2 z_3=\,0\,;\label{eq:uuddee}\\
& &-x_2\,+ \,A\,x_2^2\,\xi_3\,+\,3(x_2 y_1+x_3 y_3)\,+\,x_1 z_1\,+\,x_2 z_3=\,0\,;\nonumber\\
& &-x_3\,+ \,A\,x_3^2\,\xi_1\,\xi_2+\,3(x_1 y_3+x_4 y_3)\,+\,x_1 z_3\,+
\,x_2 z_2\,=0\,;\nonumber\\
& &-x_4\,+ \,A\,x_4^2\,\xi_1\xi_3\,+\,3(x_2 y_3+x_4 y_2)\,+\,x_3 z_3\,+\,x_4 z_2=\,0\,;\nonumber\\
& &A\,=\,\frac{m_u^2}{4\,\Lambda^2}\,\ln\frac{\Lambda^2}{\bar m^2}\,;\nonumber
\end{eqnarray}

\begin{figure}
\includegraphics[scale=0.9,width=12cm]{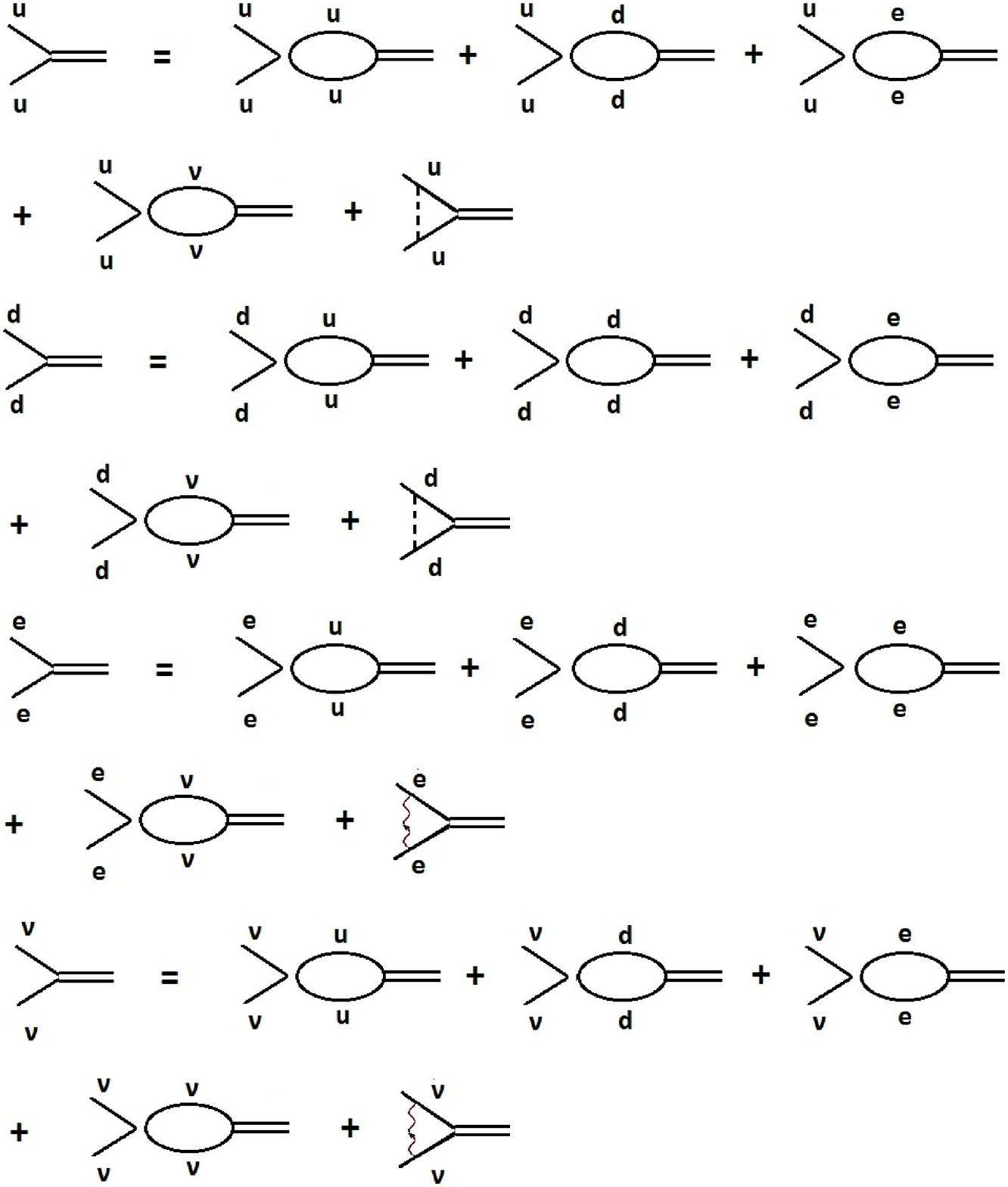}
\caption{Diagram representation of the Bethe-Salpeter
equation for scalar bound state, included in set of equations~(\ref{eq:BSUDE}).
Notations of quarks and lepton are shown by corresponding lines.
Contributions of gauge bosons exchanges (the last diagrams in each
equation) are not taken into account yet.}
\label{fig:3equat}
\end{figure}

\begin{eqnarray}
& &\nonumber\\
& &\frac{1}{B}\,=\,3(y_1\,+\,\xi_1\,y_3)\,+\,\xi_2 x_1\,+\,\xi_3 x_2;\nonumber\\
& &\frac{\xi_1}{B}\,=\,3(y_3\, + \,\xi_1\,y_2)\,+ \, \xi_2\,x_3\,+\,\xi_3 x_4;\label{eq:BSUDE}\\
& &\frac{\xi_2}{B}\,=\,3(x_1\, + \,\xi_1\,x_3)\,+ \,\xi_2\,z_1\,+\,\xi_3 z_3;\nonumber\\
& &\frac{\xi_3}{B}\,=\,3(x_2\, + \,\xi_1\,x_4)\,+ \,\xi_2\,z_3\,+\,\xi_3 z_2;\nonumber\\
& &B\,=\,1\,+\,\frac{m_0^2}{2\,\Lambda^2}\,\ln\frac{\Lambda^2}{\bar m^2}\,;
\nonumber
\end{eqnarray}
where $m_0$ is the bound state mass and $\bar m$ is an average mass of participating fermions. Let us comment the appearance of mass parameters $\xi_i$ in terms, corresponding to vertical diagrams in
Fig.~\ref{fig:6equat}. Due to the orthogonality of matrices
\begin{equation}
\frac{1+\gamma_5}{2}\,;\quad \frac{1-\gamma_5}{2}\,;\label{gamma5}
\end{equation}
terms containing $\hat q$ cancel and we are left only with mass terms in
spinor propagators. Introduction of the average $\bar m$, instead of
substituting in proper places different masses
$m_i$, means of course an approximation. However due to logarithmic
dependence on this parameter,
this approximation seems to be reasonable. Factor $A$
has to be very small and factor $B$ has to be close to unity, because
$\Lambda \gg m_i$.
Ten equations~(\ref{eq:uuddee}) correspond to the set of compensation
equations, while four equations~(\ref{eq:BSUDE}) represent the
Bethe-Salpeter equations. Let us remind, that after performing
the compensation procedure, which means exclusion of four-fermion
vertices in the newly defined free Lagrangian, we use the resulting
coupling constants in the newly defined interaction Lagrangian with the opposite sign.

The appearance of ratios $\xi_i$ in Bethe-Salpeter part~(\ref{eq:BSUDE}) of the set presumably needs explanation. We assume,
that the scalar composite state, which in our approach serves as a
substitute of the elementary Higgs scalar, consists of all existing quark-antiquark and
lepton-antilepton pairs $\bar \psi_L\,\psi_R$ (not only of heavy quarks
$\bar \Psi_L\,\Psi_R$ as in work~\cite{AZ11}). Then coupling of this
scalar with different fermions will give their masses according to
well known relation
\begin{equation}
g_a\,=\,\frac{g\,m_a}{\sqrt{2}\,M_W}\,.
\end{equation}
On the other hand, Bethe-Salpeter wave functions are proportional to coupling constants $g_a$, where $a$ is just the constituent particle. Thus we change a ratio of coupling constants by a ratio of corresponding masses $\xi_i$.

In Section~3 we consider interaction of the Higgs field also with
electroweak gauge bosons. Thus we assume, that the Higgs scalar consist of
all existing fundamental massive fields. So in future studies it should be
necessary to consider a set of Bethe-Salpeter equations including all
possible constituents. Presumably it would be advisable to take into account
also contributions of gauge interactions, which schematically presented in
triangle diagrams of Fig.~\ref{fig:3equat}.

Now let us consider solutions of set~(\ref{eq:uuddee}, \ref{eq:BSUDE}).
First of all let us remind, that parameter $A$ is very small, so we look for solutions, which are stable in the limit $A\, \to\, 0$. We also will
consider only real solutions, because our variables just correspond to
physical observable quantities.
Namely, we have for $A\,=\,0.0001$ the following real solutions
\begin{eqnarray}
& &y_1 = 0.12500,\;y_2 = y_1,\;y_3 = -\,y_1,\nonumber\\
& &z_1 = y_1,\;z_2 = y_1,\;z_3 = -\,y_1,\nonumber\\
& &x_1 = y_1,\;x_2 = -\,y_1,\; x_3 = -\,y_1,\;x_4 = y_1,\label{eq:soluden2}\\
& &\xi_1 = -\,1,\;\xi_2 = 1,\;\xi_3 = -\,1,\;B = 1.00001.\nonumber
\end{eqnarray}
\begin{eqnarray}
& &y_1 = 0.12500,\;y_2 = y_1,\;y_3 = -\,y_1,\nonumber\\
& &z_1 = y_1,\;z_2 = y_1,\;z_3 = y_1,\nonumber\\
& &x_1 = y_1,\;x_2 = y_1,\; x_3 = -\,y_1,\;x_4 = -\,y_1,\label{eq:soluden3}\\
& &\xi_1 = -\,1,\;\xi_2 = 1,\;\xi_3 = 1,\;B = 1.00001.\nonumber
\end{eqnarray}
\begin{eqnarray}
& &y_1 = 0.24999,\;y_2 = 0.33333,\;y_3 = 0,\nonumber\\
& &z_1 = 0.24999,\;z_2 = 0.56468,\;z_3 = -\,0.38570,\nonumber\\
& &x_1 = -0.24999,\, x_2 = x_3 = x_4 =0,\label{eq:soluden4}\\
& &\xi_1 = 0.86603,\;\xi_2 = -\,1,\;\xi_3 =0,\;B = 1.00003.\nonumber
\end{eqnarray}
\begin{eqnarray}
& &y_1 = 0.24999,\;y_2 = 0.33333,\;y_3 = 0,\nonumber\\
& &z_1 = 0.24999,\;z_2 = 0.99998,\;z_3 = 0,\nonumber\\
& &x_1 = -0.24999,\, x_2 = x_3 = x_4 =0,\label{eq:soluden7}\\
& &\xi_1 = 0,\;\xi_2 = 1,\;\xi_3 =0.5,\;B = 1.000025.\nonumber
\end{eqnarray}
\begin{eqnarray}
& &y_1 = 0.33332,\;y_2 = 0,\;y_3 = 0,\nonumber\\
& &z_1 = 0.24999,\;z_2 = 0.99998,\;z_3 = 0,\nonumber\\
& &x_1 =  x_2 = x_3 = x_4 =0,\label{eq:soluden8}\\
& &\xi_1 = 0,\;\xi_2 = \xi_3 = 0.57735,\;B = 1.000033.\nonumber
\end{eqnarray}
\begin{eqnarray}
& &y_1 = 0.33332,\;y_2 = 0.057288,\;y_3 = 0,\nonumber\\
& &z_1 = 0.26344,\;z_2 = 0.56470,\;z_3 = -\,0.38570,\nonumber\\
& &x_1 = x_2 = 0,\; x_3 = 0.12285,\;x_4 = -\, 0.17986,\label{eq:soluden1}\\
& &\xi_1 = \xi_2 = \xi_3 =0,\;B = 1.00003.
\nonumber
\end{eqnarray}
\begin{eqnarray}
& &y_1 = 0.29077,\;y_2 = 0.29077,\;y_3 = -\,0.04256,\nonumber\\
& &z_1 = 0.25534,\;z_2 = 0,\;z_3 = 0,\nonumber\\
& &x_1 = 0.17801,\, x_2 = x_4 =0,\,x_3 = 0.17801,\label{eq:soluden6}\\
& &\xi_1 = 1,\;\xi_2 = 1.4344,\;\xi_3 =0,\;B = 1.00003.\nonumber
\end{eqnarray}
\begin{eqnarray}
& &y_1 = 0.19313,\;y_2 = 0.18758,\;y_3 = 0.14295,\nonumber\\
& &z_1 = 0.857858,\;z_2 = 0,\;z_3 = 0,\nonumber\\
& &x_1 = -0.14116,\, x_2 =  x_4 =0,\,x_3 = 0.14393,\label{eq:soluden5}\\
& &\xi_1 = 1.069,\;\xi_2 = 0.26728,\;\xi_3 =0,\;B = 1.00002.\nonumber
\end{eqnarray}

Of course, there is a temptation to confront these solutions with
the existing generations of quarks and leptons. Let us note, that the first three solutions~(\ref{eq:soluden2},\ref{eq:soluden3},\ref{eq:soluden4}) contain mass ratios $\xi_i$ with negative signs, that is quite unnatural for fermions entering to one generation. In solutions~(\ref{eq:soluden7}, \ref{eq:soluden8}) there is no place for massless neutrino. However, these solutions may be tentatively considered in the framework of an option of wouldbe new generations with heavy neutrinos~\cite{Hill}.
For the moment, the most suitable ones are the three last
solutions~(\ref{eq:soluden1}, \ref{eq:soluden6},
\ref{eq:soluden5}).
All these solutions have nonnegative parameters $\xi_i$ and at least one lepton being massless, that might be a neutrino.
The solution~(\ref{eq:soluden1}) gives
one (the first) fundamental fermion (quark) being much heavier, than three
others, that reminds situation of the third generation
with the very
heavy $t$ quark. The solution~(\ref{eq:soluden6}) gives charged lepton mass
approximately the same
as those of quarks, that may hint the situation in the second generation
with approximately equal masses of the muon and of the $s$-quark.
The solution~(\ref{eq:soluden5}) gives
two different masses for the quark pair, while the wouldbe charged lepton
has the mass approximately four times smaller than that of the
first quark. This resembles situation for the first generation. Indeed,
let us take for the electron mass its physical value $m_e\,=\,0.51\,MeV$.
Then we have from~(\ref{eq:soluden5})

\begin{eqnarray}
& &m_e\,=\,0.51\,MeV;\nonumber\\
& &m_u\,=\,\frac{m_e}{\xi_2}\,=\,1.90\,MeV;\label{eq:mude}\\
& &m_d\,=\,\frac{m_e\,\xi_1}{\xi_2}\,=\,2.04\,MeV.\nonumber
\end{eqnarray}
The wouldbe $u$-quark mass fits into error bars of its
definition, while the wouldbe $d$-quark mass is rather lighter
than its physical value~\cite{PDG}. Note, that in our estimates we have not taken into account the phenomenon of mixing of down quarks ($d,\,s,\,b$).

Of course, the similarity is rather reluctant and there is no overall
explicit agreement with the real situation. Maybe one
could move
further with an application of a next approximation, which presumably
needs a consideration of the Bethe-Salpeter equations with account of
gauge interactions contributions, that is with account of a
gluon
exchange and of electroweak bosons exchanges. These exchanges are
schematically drawn in Fig.~\ref{fig:3equat}. The problem of an adequate formulation of the approximation needs a special
investigation. Nevertheless, even a possibility to define ratios of the fundamental masses in the compensation
approach is of a doubtless interest.

We would also draw attention to the important point, that for all
solutions parameter $B$ is close to unity, just as we have expected~\footnote
{Solutions with $B$ being not close to unity are rejected here as well 
as in what follows.}.
With decreasing of parameter $A$, which is proportional to ratio squared
of the mass of the first quark and cut-off $\Lambda$, parameter $B$ tends
to unity exactly. Emphasize, that solutions
~(\ref{eq:soluden6}, \ref{eq:soluden5}) are stable in respect to $A \to 0$.

Let us estimate also order of magnitude of mixing angles between
generations. For the purpose we introduce in effective interaction~(\ref{eq:Luuddee})  additional terms, corresponding to the
wouldbe $s,\,d$ mixing.
\begin{eqnarray}
& &\Delta\,L\,=\,\frac{8 \pi^2}{\Lambda^2}\Bigr(y_{12}\bigl(
(\bar s'_R\,d'_L+\bar d'_R\,s'_L)\bar d'_L d'_R+(\bar s'_L\,d'_R+\bar d'_L\,s'_R)\bar d'_R d'_L\bigl)+\nonumber\\
& &y_{32}\bigr((\bar s'_R\,d'_L+\bar d'_R\,s'_L)\bar s'_L s'_R+(\bar s'_L\,d'_R+\bar d'_L\,s'_R)\bar s'_R s'_L\bigl)+\label{DL}\\
& &y_{52}(\bar s'_L d'_R \bar d'_R s'_L +\bar s'_R d'_L \bar d'_L s'_R)+ t_{32}
(\bar d'_L d'_R \bar s'_L s'_R+\bar d'_R d'_L \bar s'_R s'_L)\Bigr);\nonumber
\end{eqnarray}
We have also mixing in mass terms of the two spinor fields $d',\,s'$
\begin{equation}
-m_u\bigl(\bar u u\,+\,\xi_1 \bar d' d'\,+\,\xi_4 \bar s' s'\,+\,
\xi_6(\bar s' d'\,+\, \bar d' s')\bigr)\,;\label{massmix}
\end{equation}
where, as well as in expression~(\ref{DL}), $d',\,s'$ are mixed states of
physical $d$ and $s$
\begin{equation}
d'\,=\,\cos\phi\, d\,+\,\sin\phi\,s\,;\;s'\,=\,-\,\sin\phi\, d\,+\,\cos\phi\,s\,;\label{mixCabibbo}
\end{equation}
and $\phi$ is the well known Cabibbo angle.

Now we have in addition to parameters in~(\ref{DL}) parameter $y_2$
from~(\ref{eq:GY8}), which corresponds to term $\bar d d \bar d d$ and
we also introduce the analogous parameter $y_{21}$, corresponding to term
$\bar s s \bar s s$. These variables will be fixed by
results~(\ref{eq:soluden1} - \ref{eq:soluden5}). We now neglect all other transitions but those between
$d$ and $s$ states and thus
we have the following set of equations
\begin{eqnarray}
& &-y_{12}+A y_{12}+3(y_{12} y_2+y_{32} t_{32}+2\, y_{52} y_{12})=0\,;
\nonumber\\
& &-y_{32}+A y_{32}+3(y_{12} t_{32}+y_{32} y_{21}+2\, y_{52} y_{32})=0\,;
\nonumber\\
& &-y_{52}+A y_{52}+3(y_{12}^2+y_{32}^2+2\, y_{52}^2)\,;\nonumber\\
& &-t_{32}+A t_{32}+3(y_2 t_{32}+y_{21} t_{32}+y_{12} y_{32})\,\label{setcabibbo}\\
& &\frac{\xi_1}{B}=3(y_{2}\xi_1+t_{32}\xi_4+2\, y_{12}\xi_6)\,;\nonumber\\
& &\frac{\xi_4}{B}=3(t_{32}\xi_1+y_{21}\xi_4+2\, y_{32}\xi_6)\,;\nonumber\\
& &\frac{\xi_6}{B}=3(y_{12}\xi_1+y_{32}\xi_4+2\, y_{52} \xi_6)\,.
\nonumber
\end{eqnarray}
The set has many solutions, mostly the complex ones. We consider only real solutions and choose such ones, which allow physical interpretation. Thus
we shall consider several examples and postpone for future studies the
problem of an
explanation, why just the solutions being considered correspond to real physics. Maybe this problem is connected with properties of a stability of solutions.

Fixing values for $y_2$ and $y_{21}$ from results~(\ref{eq:soluden6},  \ref{eq:soluden5}) and value $A$ we obtain seven equations for seven variables:
$y_{12},\,y_{32},\,y_{52},\,t_{32},\,B,\,\xi_1/\xi_6,\,\xi_4/\xi_6$.
Let us check if there will be a reasonable mixing of solutions~(\ref{eq:soluden6}, \ref{eq:soluden5}) that is between the first
two generations according to our guess.
With $y_2=0.18758,\,y_{21}=0.29077,\,A=0.000005,\,\xi_6=1$ we have the following solution
\begin{eqnarray}
& &y_{12}=-0.0000003158,\;y_{32}=0.078656,\;y_{52}=0.0212768,\nonumber\\
& &t_{32}=-0.000000943,\;
\xi_1=-0.0000282,\;\xi_4=3.69675,\; B=1.00003\,.\label{angleCab}
\end{eqnarray}
As well as solutions~(\ref{eq:soluden6}, \ref{eq:soluden5}), this solution
is also stable in respect to $A \to 0$.                                    
It is easy to see, that parameters $\xi_{1,4}$ give values of a mixing angle $s$ and a ratio of masses R according to the following set of equations
\begin{eqnarray}
& &s\,=\,\sin\phi\,;\quad R\,=\,\frac{m_s}{m_d}\,;\label{setangle}\\
& &(\xi_1-\xi_4)s\sqrt{1-s^2}+\xi_6(1-2 s^2)=0\,,\quad R=\frac{y+\sqrt{x^2+1}}{\sqrt{x^2+1}-y};\nonumber\\
& &x=\frac{\xi_4-\xi_1}{2},\;y=\frac{\xi_1+\xi_4}{2}\,.\nonumber
\end{eqnarray}
For data~(\ref{angleCab}) we have the following two solutions
\begin{eqnarray}
& & s_1\,=\,0.2454, \quad R_1\,=\,15.6;\label{solCab}\\
& & s_2\,=\,-\,0.9694, \quad R_2\,=\,15.6.\label{solCab2}
\end{eqnarray}
Let us note, that $s_1^2 + s_2^2 = 1$ and $R_1\,=\,R_2$ exactly.
 
Solution~(\ref{solCab}) may be compared with real situation of
$(d,\,s)$ mixing, because
mass ratio $R\,=\,m_s/m_d$ is close to its actual value
and the mixing angle is also not far from actual Cabibbo angle
value~\cite{PDG}                                                \begin{equation}                                               \sin\phi_c\,=\,s\,=\,0.2254 \pm 0.0006\,;\quad                                   \frac{m_s}{m_d}\,=\,R\,=\,19.8^{+2.4}_{-2.8}\,.\label{CabPDG}
\end{equation}

Let us try to proceed to the next approximation, that means inclusion to the 
analysis of up quarks also. This means consideration of the following effective
interaction to be added to expressions~(\ref{eq:Luuddee}, \ref{DL})
\begin{eqnarray}
& &\Delta' L = \frac{8 \pi^2}{\Lambda^2}\Bigl(
t_{21}(\bar u_L u_R \bar s'_R s'_L+\bar u_R u_L \bar s'_L s'_R) +
t_{22}(\bar u_L u_R \bar c_R c_L+\bar u_R u_L \bar c_L c_R)+ \nonumber\\
& &y_{22}\bigl(\bar u_L u_R (\bar s'_R d'_L+\bar d'_R s'_L)+
\bar u_R u_L (\bar s'_L d'_R+\bar s'_R d'_L)+h.c.\bigr)+y_{11} \bar c_L c_R
\bar c_R c_L+\nonumber\\
& &y_{21}\bar s'_L s'_R \bar s'_R s'_L+y_{42}\bigl(\bar c_L c_R
(\bar s'_R d'_L+\bar d'_R s'_L)+\bar c_R c_L
(\bar s'_L d'_R+\bar s'_R d'_L)+h.c.\bigr)+\nonumber\\
& &t_{31}(\bar s'_L s'_R \bar c_R c_L+\bar s'_R s'_L \bar c_L c_R) \Bigr)\,.
\label{cabadit}
\end{eqnarray}
Bearing in mind the stability property of solutions (\ref{eq:soluden6},
\ref{eq:soluden5}, \ref{angleCab}) in respect to $A \to 0$, we put $A =0$
(that simplifies the hunting for solutions),  and using for additional
interaction~(\ref{cabadit}) the
same rules as previously, we obtain the
following set of equations
\begin{eqnarray}
& &-y_1 + 3 (y_1^2 + y_3^2 + t_{21}^2 + t_{22}^2 + y_{22}^2) = 0;
\; -y_2  + 3 (y_2^2 + y_3^2 + t_{31}^2 + t_{32}^2 + y_{12}^2) = 0;\nonumber\\
& &-y_3+ 3 (y_3 (y_1 + y_2) + t_{21} t_{31} + t_{22} t_{32} + y_{22} y_{12});
\nonumber\\
& &-y_{12} +3 (y_{12} y_2 + y_{22} y_3 + y_{32} t_{32} + y_{42} t_{31} +
2\, y_{52} y_{12}) = 0;\nonumber\\ 
& &-y_{22}  +3 (y_{22} y_1 + y_{12} y_3 + y_{32} t_{22} + y_{42} t_{31} +
2\,y_{52}y_{22}) = 0;\nonumber\\
& &-y_{32} + 3 (y_{12} t_{32} + y_{22} t_{22} + y_{32} y_{21} +
y_{42} y_{31} +2\, y_{52} y_{32}) = 0;\nonumber\\
& &-y_{42} + 3 (y_{12} t_{31} + y_{22} t_{21} + y_{32} y_{31} +
y_{42} y_{11} +2\, y_{52} y_{42} ) = 0;\nonumber\\
& &-y_{52}+3 (y_{12}^2 + y_{22}^2 + y_{32}^ 2 + y_{42}^2 +
2\, y_{52}^2) =0  ;\label{2cabibbo}\\
& &-t_{21} + 3 (y_1 t_{21} + y_3 t_{31} + t_{21} y_{11} + t_{22} y_{31} +
y_{22} y_{42}) = 0;\nonumber\\
& &-t_{22} + 3 (y_1 t_{22} + y_3 t_{32} + t_{21} y_{31} + t_{22} y_{21}
+ y_{22} y_{32}) = 0;\nonumber\\
& &-t_{31} + 3 (y_3 t_{21} + y_2 t_{31} + t_{31} y_{11} + t_{32} y_{31} +
y_{12} y_{42}) = 0;\nonumber\\
& &-t_{32} + 3 (y_3 t_{22} + y_2 t_{32} + t_{31} y_{31} + t_{32} y_{21} +
y_{12} y_{32}) = 0;\nonumber\\
& &3 (t_{21}^2 + t_{31}^2 + 2 y_{42}^2 + y_{31}^2 +
y_{11}^2) = y_{11};\;
3 (y_{21}^2 + y_{31}^2 + 2 y_{32}^2 + t_{22}^2
+ t_{32}^2) = y_{21};\nonumber\\
& &-y_{31} + 3 (y_{11} y_{31} + y_{21} y_{31} + 2 y_{32} y_{42} +
t_{21} t_{22} + t_{31} t_{32}) = 0;\nonumber\\
& &1 = 3 B (y_1 + y_3 \xi_1 + 2 y_{22} \xi_6 + t_{21} \xi_3 + t_{22} \xi_4);\nonumber\\
& & \xi_1 = 3 B (y_3 + y2 \xi_1 + 2 y_{12} \xi_6 + t_{31} \xi_3 +
t_{32} \xi_4);\nonumber\\
& & \xi_3 = 3 B (t_{21} + t_{31} \xi_1 + y_{11} \xi_3 + y_{31} \xi_4 +
2 y_{42} \xi_6);\nonumber\\
& & \xi_4 = 3 B (t_{22} + t_{32} \xi_1 + y_{31} \xi_3 + y_{21} \xi_4 +
2 y_{32} \xi_6);\nonumber\\
& & \xi_6 = 3 B (y_{22} + y_{12} \xi_1 + y_{42} \xi_3 + y_{32} \xi_4 +
2\,y_{52}  \xi_6).\nonumber
\end{eqnarray}
Here $B$, which has to be equal to unity, is the same as in~(\ref{setcabibbo}).
Additional mass parameters are defined in the following way by extending
~(\ref{massmix}) to the following expression
\begin{equation}
-m_u\bigl(\bar u u\,+\,\xi_1 \bar d' d'\,+\,\xi_3 \bar c c\,+\,\xi_4 \bar s' s'\,+\,
\xi_6(\bar s' d'\,+\, \bar d' s')\bigr)\,;\label{massmix1}
\end{equation}
There is a solution of set~(\ref{2cabibbo}), which is close to previous
one~(\ref{solCab}). Namely it looks like for A=0
\begin{eqnarray}
& &y_1=0.1773,\;y_2=0.1571,\;y_3=0.16583,\;y_{11}=0.3329,\;
y_{21}=0.3327,\nonumber\\
& &y_{31}=0.00052098,\;y_{12}=y_{22}=y_{32}=y_{42}=0,\;
y_{52}=0.166667,\label{sol2cab}\\
& &t_{21}=0.0082035,\;t_{22}=-0.0099095,\;t_{31}=-0.0087183,\;
t_{32}=0.010531,\;\nonumber\\
& &\xi_1=1.190304,\;\xi_3=9.97278,\;\xi_4=12.42852,\,\xi_6=2.68897.\nonumber
\end{eqnarray}
Solution~(\ref{sol2cab}) gives the following results for
parameters~(\ref{setangle})
\begin{equation}
s\,=\,0.221,\quad R\,=\,22.43.\label{Cabibbo22}
\end{equation}
We see, that this result agrees actual values~(\ref{CabPDG}) even better
than result~(\ref{solCab}). That is we may state the improvement  of results
in the course of successive approximations.

As a matter of fact solution~(\ref{sol2cab}) gives the wouldbe $c$-quark mass
only ten times more than that of the $u$-quark. However, one may expect strong influence on this relation of a mixing with the heavy $t$-quark. Thus the
approximation, which we demonstrate here is applied just for consideration of
the $d\,s$ mixing.

The examples being just considered shows possibility of definition of
mass ratios and of some mixing angles in the compensation approach.
There are also other mixing angles
in the Standard Model, first of all, the famous Weinberg angle $\theta_W$ in
$W^0,\,B$ mixing. In  the next section we consider a possible way of
calculation of this important parameter following the same approach.

\section{Weinberg mixing angle and the fine structure constant}

Let us demonstrate a simple model, which illustrates how the well-known Weinberg mixing angle
could be defined. Let us  consider a possibility of a spontaneous generation of the following
effective interaction of electroweak gauge bosons
\begin{eqnarray}
& & L_{eff}^{W} = G_1 W^a_\mu W^d_\mu\, W^a_{\rho \sigma} W^d_{\rho \sigma} +
G_2 W^a_\mu W^a_\mu W^b_{\rho \sigma} W^b_{\rho \sigma} +\nonumber\\
& & G_3\,W^a_\mu W^a_\mu B_{\rho \sigma} B_{\rho \sigma} +
G_4\,Z_\mu Z_\mu W^b_{\rho \sigma} W^b_{\rho \sigma} + 
G_5\,Z_\mu Z_\mu B_{\rho \sigma} B_{\rho \sigma} .\label{eq:LeffWZ}
\end{eqnarray}
where we maintain the residual gauge invariance for the electromagnetic
field. Here indices $a,d$ correspond to charged $W$-s, that is they take
values $1,\,2$, while index $b$
corresponds to three components of $W$ defined by the initial formulation
of the electro-weak interaction. Let us remind the relation, which connect fields $W^0,\,B$ with physical fields of the $Z$ boson
and of the photon
\begin{eqnarray}
& &W^0_\mu = \cos \theta_W\,Z_\mu + \sin \theta_W\,A_\mu ;\nonumber\\
& &B_\mu = -\,\sin \theta_W\,Z_\mu + \cos \theta_W\,A_\mu .
\label{eq:Wmix8}
\end{eqnarray}
Interactions of type~(\ref{eq:LeffWZ}) were earlier introduced on phenomenological
grounds in works~\cite{BB1,BB2}.
Let us introduce an effective cut-off $\Lambda$ in the same way as we
have done in the previous section and use for definition of $\Lambda$ relation~(\ref{eq:Lambda}).
Here we shall proceed just in the same way as earlier. Then let us consider a possibility of a spontaneous generation of interaction~(\ref{eq:LeffWZ}).
In doing this we again proceed with the
add-subtract procedure, which was used throughout works~\cite{BAA04}\cdash\cite{AVZ2}. Now we start with
usual form of the Lagrangian, which describes electro-weak gauge
fields $W^a$ and $B$
\begin{eqnarray}
& &L = L_0\,+\,L_{int}\,;\nonumber\\
& &L_0 = -\,\frac{1}{4}\bigl( W_{0\mu\nu}^a\,W_{0\mu\nu}^a\bigr)\,-\,\frac{1}{4}\bigl( B_{\mu\nu}\,B_{\mu\nu}\bigr)\,;\label{eq:L0WB}\\
& &L_{int}\,=\,-\frac{1}{4}\bigl( W_{\mu\nu}^a\,W_{\mu\nu}^a-W_{0\mu\nu}^a\,W_{0\mu\nu}^a\bigr) \,.\label{eq:LIntWB}\\
& &W_{0\mu\nu}^a = \partial_\mu W_\nu^a - \partial_\nu W_\mu^a;\;
B_{\mu\nu} = \partial_\mu B_\nu - \partial_\nu B_\mu .\nonumber
\end{eqnarray}
and $W_{\mu\nu}^a$ is the well-known non-linear Yang-Mills field of
$W$-bosons.
Then we perform the add-subtract procedure of expression~(\ref{eq:LeffWZ})
\begin{eqnarray}
& &L = L'_0\,+\,L'_{int}\,;\nonumber\\
& &L'_0 = L_0\,-\,L_{eff}^{W}\,; \label{eq:L'WB0}\\
& &L'_{int}\,=\,L_{int}\,+\,L_{eff}^{W}\,. \label{eq:L'WBInt}
\end{eqnarray}

Now let us formulate compensation equations.
We are to demand, that considering the theory with Lagrangian
$L'_0$~(\ref{eq:L'WB0}), all contributions to four-boson
connected vertices, corresponding to interaction~(\ref{eq:LeffWZ}) are summed up to zero. That is the undesirable interaction part in the would-be free Lagrangian~(\ref{eq:L'WB0}) is compensated. Then we are rested with
interaction~(\ref{eq:LeffWZ}) only in the proper place~(\ref{eq:L'WBInt})
We have the following set of compensation equations, which corresponds to
diagrams being presented in the first six rows of Fig.~\ref{fig:NJLBS}
\begin{eqnarray}
& & -\,x_1\,+\,x_1^2\,=\,0\,;\nonumber\\
& &-\,x_2\, +\,2\,x_2^2\,+\,2\,x_1 x_2\,+\,(1-a^2)\,x_3  x_4\,+\nonumber\\
& &a^2\,x_2 x_4=\,0\,;\nonumber\\
& & -\,x_3\,+\,x_1 x_3\,+\,2\,x_2 x_3\,+\,a^2\,x_2\,x_5\,+\nonumber\\
& &(1-a^2)\,x_3 x_5\,=\,0\,;\label{eq:CompWB}\\
& & -\,x_4\,+\,x_1 x_4\,+\,2\,x_2 x_4\,+\,a^2\,x_4 x_5\,=\,0\,;\nonumber\\
& &-\,x_5\,+\,2\,x_3 x_4\,+\,a^2\,x_4\,x_5\,+\,(1-a^2)\,x_5^2\,=\,0\,;\nonumber\\
& &x_i\,=\,\frac{3\,G_i\,\Lambda^2}{64\,\pi^2}\,;\quad a\,=\,\cos\theta_W\,.\nonumber
\end{eqnarray}
Factor $2$ in several terms of equations here corresponds to sum by weak
isotopic index $\delta^a_a\,=\,2,\,a = 1,\,2$.
\begin{figure}
\includegraphics[scale=0.3]{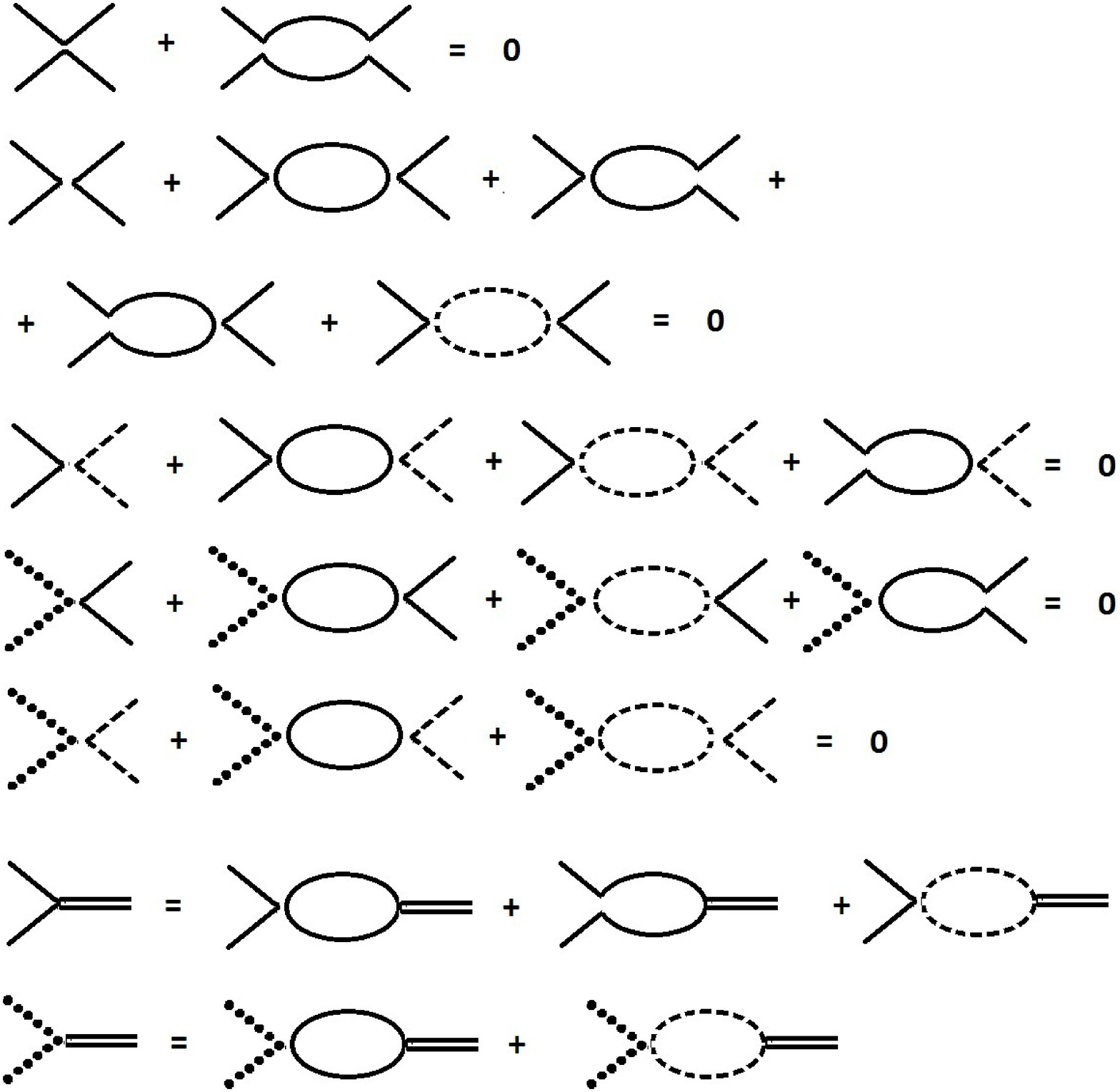}
\caption{Diagram representation of set~(\ref{eq:CompWB})
(the first five equations) and~(\ref{eq:BS}) (the two last ones).
Simple line represent
$W$-s, dotted lines represent $B$ and lines, consisting of black spots, represent $Z$. Double lines represent the Higgs scalar.}
\label{fig:NJLBS}
\end{figure}

Then, following the reasoning of the approach, we assume, that the Higgs scalar corresponds to a bound state consisting  of a complete set of fundamental particles. Note, that in  work~\cite{AZ11} we have considered only the heaviest particle $t$ quark as the main constituent of the Higgs scalar. Here we
are to include the electro-weak bosons.
There are two Bethe-Salpeter equations for this bound state, because constituents are either $W^a\,W^a$ or $Z\,Z$. These equations are presented in the last two rows of Fig.~\ref{fig:NJLBS}. In approximation of very large cut-off $\Lambda$ these equations have the following form
\begin{eqnarray}
& &x_1\,+(2\,+\,a)\,x_2\,+\,\frac{1-a^2}{a}\,x_3\,+\,\beta\,=\,1\,;
\label{eq:BS}\\
& &(2\,+\,a)\,x_4\,+\,\frac{1-a^2}{a}\,x_5\,+\,\frac{\beta}{a}\,=
\,\frac{1}{a}\,.\nonumber
\end{eqnarray}
Here we introduce parameter $\beta$, which describes wouldbe additional contributions. We consider as physical solutions those with very small $\beta$.
Now we look for solutions of set~(\ref{eq:CompWB}, \ref{eq:BS}) for
variables $x_i,\,a,\,\beta$.
Of course, there is the
trivial solution: all $x_i\,=\,0,\,\beta\, =\, 1$. However there are also non-trivial solutions. Namely, there are the the following two ones with $x_1\,=\,1$
\begin{eqnarray}
& &x_2\,=\,0\,;\; x_3\,=\,0.729625\,;\; x_4\,=\,0\,;\; x_5\,=\,0\,;\label{eq:Sol1WZ}\\
& &\beta_1\,=\,1\,;\;\beta_2\,=\,\frac{0.729625\,(a-1)}{a}\,;\nonumber
\end{eqnarray}
for any $a$,
and the following three ones with $x_1\,=\,0$
\begin{eqnarray}
& &x_2\,=\,0\,,\;x_3\,=\,3.070337\,,\;x_4\,=\,0\,,\; x_5\,=\,3.61378\,,\nonumber\\
& &a\,=\;0.8504594\,,\;\beta \,=\, -\,5.06\cdot 10^{-16}\,;
\label{eq:Sol0WZ}\\
& &x_2\,=\,0.48772\,,\;x_3\,=\,0\,,\;x_4\,=\,1.2654\,,\; x_5\,=\,0\,,\nonumber\\
& &a\,=\,0.33801\,,\;\beta \,=\, -\,1.2\cdot 10^{-5}\,;\nonumber\\
& &x_2\,=\,0.5\,,\;x_3\,=\,1.09555\,,\;x_4\,=\,0\,,\; x_5\,=\,0\,,\nonumber\\
& &a\,=\,-\,0.75556\,,\;\beta =\,1\,.\nonumber
\end{eqnarray}
Very small $\beta$ are appropriate for
the first solution of~(\ref{eq:Sol0WZ}) with
$\beta \simeq -\,5\cdot 10^{-16}$ and for the second one with
$\beta \simeq -\,1.2\cdot 10^{-5}$. Note, that for solutions~(\ref{eq:Sol1WZ})
smallness of $\beta$ is achieved only for the second one with $a\,\to\,1$, that is in an absence of the mixing.
The solution with the smallest $\beta$ gives for the mixing parameter
\begin{equation}
\sin^2\theta_W\,=\,1-a^2\,=\,0.27672\,.\label{eq:sintheta0}
\end{equation}
This value corresponds to scale $\Lambda$~(\ref{eq:Lambda}), which
is defined by parameter $z_0$.
At this scale the electroweak coupling according to~(\ref{eq:gz0}) is the
following
\begin{equation}
\alpha_{ew}(z_0)\,=\,\frac{g(z_0)^2}{4\,\pi}\,=\,0.028999\,.
\label{eq:aeW}
\end{equation}
Then we obtain the electromagnetic coupling at the same scale
\begin{equation}
\alpha(z_0)\,=\,\alpha_{ew}(z_0)\,\sin^2\theta_W(z_0)\,=\,0.0080244\,.
\label{eq:aelectr}
\end{equation}
With the well-known evolution expression for electromagnetic coupling we
have for six quark flavors ($\Lambda \gg M_W$)
{\large
\begin{equation}
\alpha(z_0)\,=\,\frac{\alpha(M_Z)}{1-\frac{5 \alpha(M_Z)}{6\,\pi}\ln\bigl[\frac{\Lambda^2}{M_Z^2}]}\,=\,0.0080244\,.
\end{equation}
}
This gives for value $\Lambda$ from expression~(\ref{eq:Lambda}) with an account of~(\ref{eq:gz0})
\begin{equation}
\alpha(M_Z)\,=\,0.007719\,.\label{eq:alpha}
\end{equation}
to be compared with experimental value~\cite{PDG}
\begin{equation}
\alpha(M_Z)\,=\,0.0077562\pm0.0000012.\label{eq:alphaexp}
\end{equation}
Of course, set of equations~(\ref{eq:CompWB}, \ref{eq:BS}) is approximate.
It quite may be, that with an account of necessary corrections the agreement
of the result with experimental number~(\ref{eq:alphaexp})
will be not such indecently good. For example, provided we take the value of
boundary momentum $\Lambda$ being an order of magnitude up and down of that defined by relations~(\ref{eq:gz0}), we have
\begin{equation}
\alpha(M_Z)_{up}\,=\,0.00765\,;\quad \alpha(M_Z)_{down}\,=\,0.00779\,.
\label{eq:alphaud}
\end{equation}
The second solution gives mach larger value for
$\sin^2\theta_W \simeq 0.89$. As a result this leads to
$\alpha(M_Z)\simeq 0.0235$, that is three times more, than~(\ref{eq:alpha}, \ref{eq:alphaexp}).
Now we have one solution~(\ref{eq:alpha}) being in agreement with actual physics and another one being in evident disagreement. Which one is to be
used?

The answer is connected with the problem of a stability of solutions~(\ref{eq:Sol0WZ}).
The stability in the model is defined by sum of vacuum averages
\begin{equation}
\frac{1}{4}<W_{\mu \nu}^a\,W_{\mu \nu}^a>\,+\,\frac{1}{4}<B_{\mu \nu}\,
B_{\mu \nu}>\,.\label{eq:VEVWB}
\end{equation}
A calculation of these vacuum averages even in the first approximation
needs knowledge of explicit form-factors in effective interactions~(\ref{eq:LeffWZ}).
To achieve this knowledge one has to perform the next step in
a formulation and a solution of compensation equations, namely, it is
necessary to take into account two-loop
terms in compensation equations in analogy to
works~\cite{BAA06,AVZ06}. This procedure is to be considered elsewhere.
For the moment we may only state, that one of two possible solutions gives
satisfactory value for fine structure constant $\alpha(M_W)$.
On the other hand, let us note the following. Provided the form-factor will be qualitatively the same as is presented in Fig.~\ref{fig:compenG}, {\it i.e.}
being negative for large momenta, preliminary estimates show, that just the
solution with value $\alpha(M_W)$~(\ref{eq:alpha}) is more stable than
other one.
Maybe it is worth mentioning, that the preferable solution contains only combination
$B_{\mu\nu}B_{\mu\nu}$ in effective interaction~(\ref{eq:LeffWZ}), while the solution with large $\alpha(M_W)$ on the contrary contains only combination
$W^b_{\mu\nu}W^b_{\mu\nu}$.

The results being demonstrated can not be regarded as
finally decisive ones
and are rather indications of how things might occur. However in view of
a fundamental importance of a possibility to define parameters of the
Standard Model, we do present these considerations. Additional arguments
on behalf of our point of view are presented in the subsequent section.

\section{Conclusion}
Possible way of determination of fundamental fermion mass ratios,
of mixing angles in the Cabibbo-Kobayashi-Maskawa matrix and of the
Weinberg mixing angle, which is
proposed in the work needs further studies, especially in respect to
the next approximations. As well
problems of stability, which might choose appropriate solutions,
need thorough consideration. Thus we can not consider results being
described here as final ones. They are just examples, which illustrate how things may occur.

In any case the examples being considered in the present work show,
that a consideration of
effective interactions in the
compensation approach might lead to a determination of fundamental
parameters of the Standard Model
including the Weinberg mixing angle, mass ratios of fundamental particles
and the Cabibbo angle. Remind, that a
result being obtained above give quite a satisfactory value for the most important physical
parameter -- the fine structure constant $\alpha$. We would also draw attention to an appearance of very small numbers in solutions being considered.
{\it E.g.} solution~(\ref{eq:Sol0WZ}) contains parameter
$\beta \simeq 5\cdot 10^{-16}$. This might be useful in application to problems of
hierarchy~\cite{Gildener,Witten}.

Let us emphasize, that the possibility of an adequate definition of the
fundamental parameters of the Standard Model, is alternative to the
option of
anthropic principle
(see recent works and reviews~\cite{Hogan}\cdash\cite{Meissner} and papers quoted therein), which assumes
multiplicity of Universes. The main foundation of this postulate is just
an absence of
any mechanism, which could fix values of parameters of the Standard Model.
The number $N_{SM}$ of fundamental parameters of the Standard Model including those, which are related to neutrinos, may be estimated to be as large as
25. Because each possible set of these parameters corresponds to a really existing Universe, the power of the set of the totality of Universes corresponds to the continuum. 
On the other hand, the existence of a human being, who is capable to observe the Nature and to try to understand Its laws, is closely connected with actual values of the parameters of the Standard Model.
The properties of nuclei are connected with parameters defining low-energy strong interaction, that is the average strong coupling
at low energies
$\bar \alpha_s$ and light quark masses $m_u,\,m_d$. The most important parameters,
which define the rich variety of organic substances, which is inevitably necessary for the life
generation and evolution, are just the fine structure constant $\alpha$ and the electron mass $m_e$. We have discussed in the present work possibilities for determination of all these fundamental parameters, but strong coupling
$\bar \alpha_s$, which was considered in work~\cite{AZ20132}.

Thus the anthropic principle
assumes, that we live in the only Universe, which supplies conditions for an existence of a human being, that is in the Universe with such parameters $\alpha,\,\bar \alpha_s,\,m_u,\,m_d,\,m_e$, which we consider now
as real physical ones. All other Universes are deprived of an observer
and so are principally unobservable.

The approach, which we have used in the present work, provides a possibility to define at least some of these parameters. Indeed, in work~\cite{AZ20132} we have obtained value of average strong coupling in the low-momenta region
$\bar \alpha_s \simeq 0.85$  in agreement with its phenomenological value.
As for other parameters, in the present work we just discuss examples of definition of the fine structure constant and light mass ratios in the framework of a spontaneous generation
of effective interactions in the
Standard Model. Relations~(\ref{eq:gz0}, \ref{eq:mude}, \ref{solCab}, \ref{Cabibbo22},
 \ref{eq:alpha}) seemingly can
not be yet
considered being decisive ones, but the examples, which give these results, may serve as leading indications for further more detailed studies. In case
of a realization of the program, we would obtain an
understanding of how
values of the fundamental parameters are fixed. Then the conception of the uniqueness of the Universe
might be established. That is, it might be, that the observable Universe corresponds to the most stable
non-trivial solution of the Standard Model. The authors do express the conviction, that
a possible way to this goal is connected with a phenomenon of a
spontaneous generation of
effective interactions in the framework of the Standard Model.

\section{Acknowledgments}

The work is supported in part by the Russian Ministry of Education and Science
under grant NSh-3042.2014.2.


\begin{thebibliography}{**}
\bibitem{BAA04} B. A. Arbuzov, Theor. Math. Phys., {\bf 140}, 1205 (2004);
\bibitem{BAA06}B. A. Arbuzov, Phys. Atom. Nucl., {\bf 69}, 1588 (2006).
\bibitem{AVZ06}B. A. Arbuzov, M. K. Volkov and I. V. Zaitsev, Int. J. Mod.
Phys. A, {\bf 21}, 5721 (2006).
\bibitem{BAA09}B. A. Arbuzov, Eur. Phys. J., {\bf C61}, 51 (2009).
\bibitem{AZ11} B. A. Arbuzov and I. V. Zaitsev, Int. J. Mod. Phys.,
{\bf A26}, 4945 (2011).
\bibitem{AVZ2} B. A. Arbuzov and I. V. Zaitsev,  	
 Phys. Rev., {\bf D85}: 093001 (2012).
\bibitem{AZ20132} B. A. Arbuzov and I.V. Zaitsev, Int. J. Mod. Phys.,
{\bf A28}: 1350127 (2013).
\bibitem{Bog1} N. N. Bogoliubov, Soviet Phys.-Uspekhi, {\bf 67}, 236 (1959).
\bibitem{Bog2} N. N. Bogoliubov, Physica Suppl. (Amsterdam), {\bf 26}, 1 (1960).
\bibitem{ABOOK} B. A. Arbuzov, {\it Non-perturbative Effective Interactions
in the Standard Model}, De Gruyter, Berlin, 2014.
\bibitem{Hag1} K. Hagiwara, R. D. Peccei, D. Zeppenfeld and K. Hikasa,
Nucl. Phys., {\bf B282}, 253 (1987).
\bibitem{Hag2} K. Hagiwara, S. Ishihara, R. Szalapski and D. Zeppenfeld,
Phys. Rev., {\bf D48}, 2182 (1993).
\bibitem{PDG} K. A. Olive {\it et al.} (Particle Data Group),
{\it Review of particle physics}, Chin. Phys. {\bf C38}: 090001 (2014).
\bibitem{Nambu} Y. Nambu and G. Jona-Lasinio, Phys. Rev., {\bf 122}, 345 (1961).
\bibitem{Nambu2} Y. Nambu and G. Jona-Lasinio, Phys. Rev., {\bf 124}, 246 (1961).
\bibitem{Jap} T. Eguchi, Phys. Rev., {\bf D14}, 2755 (1976).
\bibitem{ERV} D. Ebert, H. Reinhardt and M. K. Volkov, Prog. Part. Nucl.
Phys., {\bf 33}, 1 (1994).
\bibitem{VolRad} M. K. Volkov and A. E. Radzhabov, Phys. Usp., {\bf 49},
551 (2006).
\bibitem{Hill} C. T. Hill and E.A. Paschos, Phys. Lett. {\bf B241},
96 (1990).
\bibitem{BB1} G. Belanger and F. Boudjema, Phys. Lett., {\bf B288}, 201
(1992).
\bibitem{BB2} G. Belanger {\it et al.}, Eur. Phys. J., {\bf C13}, 283
(2000).                                                                                                                                                                                                                                                                                                                                                                                                                                                                       \bibitem{Gildener} E. Gildener, Phys. Rev., {\bf D14}, 1667 (1976).
\bibitem{Witten} E. Witten,  Phys. Lett., {\bf B105}, 267 (1981).
\bibitem{Hogan} C. J. Hogan, Rev. Mod. Phys., {\bf 72}, 1149 (2000).
\bibitem{Jaffe} R. L. Jaffe, A. Jenkins and I. Kimchi, Phys. Rev., {\bf D79}: 
065014 (2009).
\bibitem{Schell} A. N. Schellekens, Rev. Mod. Phys., {\bf 85}, 1491 (2013). 
\bibitem{Meissner} U.-G. Meissner, arXiv: 1409.2959 (2014). 
\end{thebibliography}
\end{document}